\documentclass[twocolumn,showpacs,preprintnumbers,amsmath,amssymb,prl,superscriptaddress]{revtex4}

\usepackage{graphicx}
\usepackage{dcolumn}
\usepackage{bm}

\begin{document}

\title{Pristine and Intercalated  Transition Metal Dichalcogenide Superconductors} 
\author{Richard A. Klemm}
\affiliation{Department of Physics, University of Central Florida, Orlando, Florida 32816-2385, USA,\\
email: richard.klemm@ucf.edu, telephone: (+1) 407-882-1160, FAX: (+1) 407-823-5112 }
\date{\today}

\begin{abstract}
Transition metal dichalcogenides (TMDs) are quasi-two-dimensional layered compounds that exhibit strongly competing effects of charge-density wave (CDW) formation and superconductivity (SC).  The weak van der Waals interlayer bonding between hexagonal layers of octahedral or trigonal prismatic TMD building blocks allows many polytypes to form.  In the single layer $1T$ polytype materials, one or more  CDW states can form, but the pristine TMDs are not superconducting. The  $2H$ polytypes have  two or more Fermi surfaces and saddle bands, allowing for dual orderings, which can be coexisting CDW and SC orderings, two SC gaps as in MgB$_2$,  two CDW gaps,  and possibly even pseudogaps above the onset $T_{\rm CDW}$s of CDW orderings.  Higher order polytypes allow for multiple CDW gaps and at least one superconducting gap.  The CDW transitions $T_{\rm CDW}$s usually greatly exceed the superconducting transitions at their low $T_{\rm c}$ values, their orbital order parameters (OPs) are generally highly anisotropic and can even contain nodes, and the SC OPs can be greatly affected by their simultaneous presence. The properties of the CDWs ubiquitously seen in TMDs are remarkably similar to those of the pseudogaps seen in the high-$T_{\rm c}$ cuprates. In 2$H$-NbSe$_2$, for example, the CDW renders its general $s$-wave SC OP orbital symmetry to be highly anisotropic and strongly reduces its Josephson coupling strength ($I_{\rm c}R_{\rm n}$) with the conventional SC, Pb.  Hence, the pristine TMDs are highly ``unconventional'' in comparison with Pb, but are much more ``conventional'' than are the  ferromagnetic superconductors such as URhGe.   Applied pressure and intercalation generally suppress the TMD CDWs, allowing for enhanced SC formation, even in the $1T$ polytype materials.  The misfit intercalation compound (LaSe)$_{1.14}$(NbSe$_2$) and many $2H$-TMDs intercalated with organic Lewis base molecules, such as TaS$_2$(pyridine)$_{1/2}$, have completely incoherent $c$-axis transport,  dimensional-crossover effects, and behave as stacks of intrinsic Josephson junctions.  Except for the  anomalously large apparent violation of the Pauli limit of the upper critical field of (LaSe)$_{1.14}$(NbSe$_2$), these  normal state and superconducting properties of these intercalation compounds are very similar to those seen in the high-$T_{\rm c}$ superconductor, Bi$_2$Sr$_2$CaCu$_2$O$_{8+\delta}$ and in the organic layered superconductor, $\kappa$-(ET)$_2$Cu[N(CN)$_2$]Br, where ET is bis(ethylenedithio)tetrathiafulvalene.  Electrolytic intercalation of TMDs with water and metallic ions leads to compounds with very similar properties to cobaltates such as Na$_x$CoO$_2\cdot y$H$_2$O.
\end{abstract}

\maketitle


\section{Introduction}
\begin{figure}
\center{\includegraphics[width=0.15\textwidth]{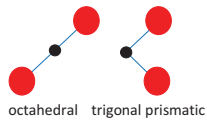}
\caption{(color online) Sketches of the two TMD configurations (or building blocks).  Left:  octahedral.  Right:  trigonal prismatic.  The small black and large red filled circles respectively represent the transition metal and chalcogen ions.}}\label{fig1}
\end{figure}
The transition metal dichalcogenides (TMDs) are a very interesting class of layered superconductors that  exhibit many features  strikingly similar to those seen in the high-transition temperature $T_c$ cuprates,  MgB$_2$, the organic layered superconductors, the cobaltates, and  the iron-based pnictides \cite{Klemm1,Klemm2}.  Pristine TMDs are constructed from either the octahedral or trigonal prismatic topologically inequivalent T$\chi_2$ building blocks, as sketched in Fig. 1, where T is a group IV, V, or VI transition metal and $\chi$ is a chalcogen, either S, Se, or Te\cite{Klemm1,WDM}.  Because of the small differences in formation energies, the TMDs form in many polytypes. The most common polytypes of TiSe$_2$, NbSe$_2$, and MoS$_2$ are respectively the 1$T$,  2$H$(a), and 2$H$(b) forms, which respectively consist of identical octahedral layers and two types of trigonal prismatic double layers. TaS$_2$ also exists in many other polytypes consisting of ordered one-dimensional arrays of these two building blocks.  With only a very few exceptions, the TMDs exhibit one or more charge-density waves (CDWs) \cite{WDM}, which can even exhibit nodes in their energy gaps \cite{Liu}. One TMD is presently known not to exhibit any CDWs, but instead exhibits two distinct superconducting energy gaps.  These features occur in the absence of any observable magnetism. Angle-resolved photoemission spectroscopy and scanning tunneling microscopy have revealed that the CDW and superconducting orderings coexist, and compete for Fermi surface and/or saddle band gap formation area in pristine TMDs.  Upon intercalation with a large variety of elements or compounds, the CDWs are removed, and the resulting superconductors are often extremely anisotropic in their normal and superconducting state properties, even exhibiting clear signs of dimensional-crossover effects.

\section{Pristine Transition Metal Dichalcogenides}

The TMDs are produced in vapor phase transport reactions, by adding either a small amount of $I_2$ or excess chalcogen compositions to the stochiometric elemental mix,  to facilitate the vapor phase transport in a two- or three-zone furnace \cite{Schaefer}.  The electronic properties of layers of the two $T\chi_2$ configurations are generally very different.   In all known examples of the 1$T$ polytype, such as $1T$-TaS$_2$, $1T$-TiS$_2$, $1T$-TiSe$_2$, and $1T$-VSe$_2$, the pristine materials were never found to be superconducting at ambient pressure \cite{Thompson1,Sipos,Nohara}, and the normal state transport properties were found to be either semiconducting or semi-metallic within the layers at temperatures $T$ approaching or above room temperature.  This non-metallic behavior is due to the formation of a CDW that opens up a gap or pseudogap over most of the surprisingly complex Fermi surface for  $T$ in the vicinity of room temperature \cite{Thompson1,WDM,Zhang}, resulting in the complex set of transitions apparent in the in-plane resistivity $\rho_{||}(T)$ of $1T$-TaS$_2$ shown in Fig. 2.
\begin{figure}
\center{\includegraphics[width=0.4\textwidth]{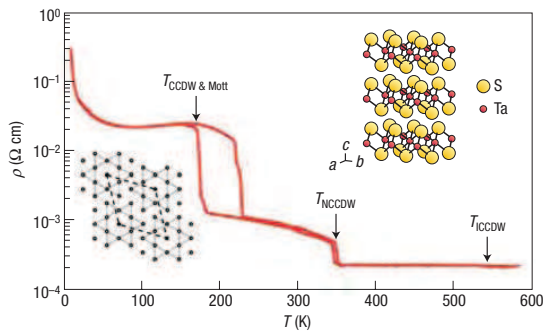}
\caption{(color online) Logarithmic plot of the in-plane resistivity $\rho_{||}(T)$ of 1$T$-TaS$_2$. Reprinted with permission of B. Sipos, A.F. Kusmartseva, A. Akrap, H. Berger, L. Forr{\'o}, E. Tuti{\v s}. From Mott state to superconductivity in $1T$-TaS$_2$. Nature Mat. 7 (2008) 960. Copyright \copyright2008 Nature Publishing Group.}}\label{fig2}
\end{figure}
With about 2 GPa of applied pressure, $1T$-TaS$_2$ becomes superconducting, and with increased pressure, both the Mott phase and nearly commensurate CDW are suppressed, and  superconductivity persists up to the transition temperature $T_c$~$\sim$~5 K for pressures up to 25 GPa \cite{Sipos}.

\begin{figure}
\center{\includegraphics[width=0.25\textwidth]{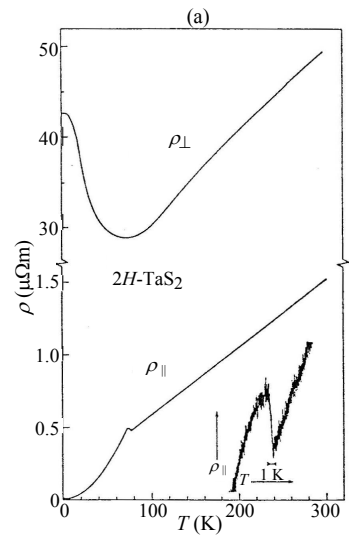}
\caption{Plots of the normal state resistivities $\rho_{||}(T)$ and $\rho_{\perp}(T)$ of 2$H$-TaS$_2$. The inset is an enlargement of the  weakly commensurate CDW transition at 75 K in $\rho_{||}(T)$.  Reprinted with permission of J.P. Tidman, O. Singh, A.E. Curzon, R.F. Frindt. The phase transition  in 2$H$-TaS$_2$ at 75 K.  Phil. Mag. 30 (1974) 1191. Copyright \copyright1974.  Taylor \& Francis.}}\label{fig3}
\end{figure}
In the $2H$-TMDs, the two layers are in one of two orientations of trigonal prismatic $T\chi_2$ configurations (Fig. 1), and the normal state behavior is considerably more varied than in the $1T$-TMDs.
 In $2H$(a)-TaS$_2$, usually written as 2$H$-TaS$_2$, there is only one CDW transition at $T_{\rm CDW}=75$ K, which is weakly first-order, as indicated resistively  in Fig. 3 \cite{Tidman}.  In this case, the CDW opens up a gap on the saddle-shaped hole band below $T_{\rm CDW}$, changing the Hall constant from positive to negative, \cite{TGK,NaitoTanaka},  and the $c$-axis resistivity $\rho_{\perp}(T)$ has a minimum followed by a rise as $T$ decreases down to $T_{\rm c}$~$\sim$~0.6 K \cite{Tidman}, where the compound goes superconducting at ambient pressure \cite{Tsutsumi,Garoche2}.  Meanwhile  the magnetic  susceptibility $\chi_{\perp}(T)$ for ${\bm H}||\hat{\bm c}$ decreases sharply and continuously below $T_{\rm CDW}$  \cite{WDM}. Since no long-range magnetic order was ever detected in this material, this continuous decrease in $\chi_{\perp}(T)$ below $T_{\rm CDW}$ can presumably be attributed to the decrease in the density of states on the hole band due to the opening of the CDW gap \cite{WDM,TGK}. The first Fermi surface measurements on $2H$-TaS$_2$ were made using de Haas Shubnikov and de Haas van Alphen oscillations, indicating that the Fermi surface is not as 2D as is that of 4$H$(b)-TaS$_2$ \cite{Hillenius}. Interestingly, it was shown using angle-resolved photoemission spectroscopy (ARPES) \cite{Tonjes}, that the CDW gap in 2$H$-TaS$_2$ actually appears to exhibit a node on the Fermi surface, as detailed in Fig. 4.
\begin{figure}
\center{\includegraphics[width=0.45\textwidth]{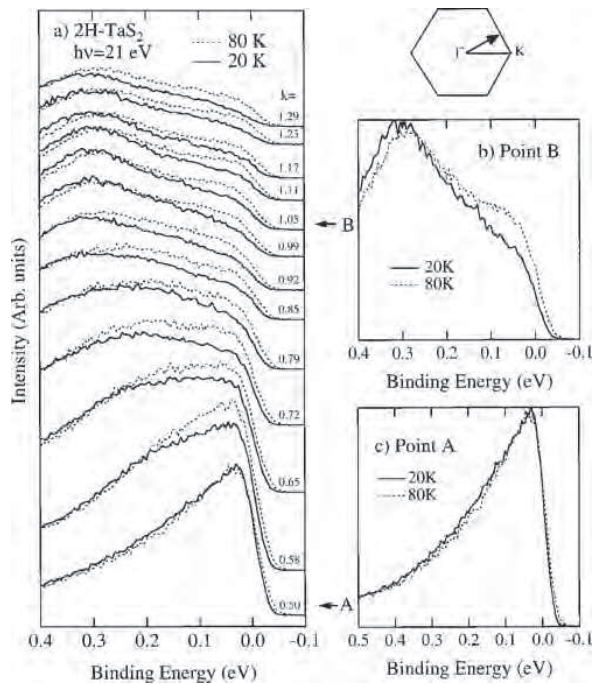}
\caption{ARPES study at 80 K (dashed) and 20 K (solid) of the CDW gap anisotropy in 2$H$-TaS$_2$.  The energy dispersion curves are directed 36$^{\circ}$ above the $\Gamma-K$ line, and the enlargement  details the vanishing of the CDW gap at the nodal point $A$.  Reprinted with permission of W.C. Tonjes, V.A. Greanya, R. Liu, C.G. Olson, P. Molini{\'e}. Charge-density wave mechanism in the 2$H$-NbSe$_2$ family:  Angle-resolved photoemission study.  Phys. Rev. B 63 (2001) 235101.  Copyright \copyright2001, American Physical Society. }}\label{fig4}
\end{figure}

\begin{figure}
\center{\includegraphics[width=0.4\textwidth]{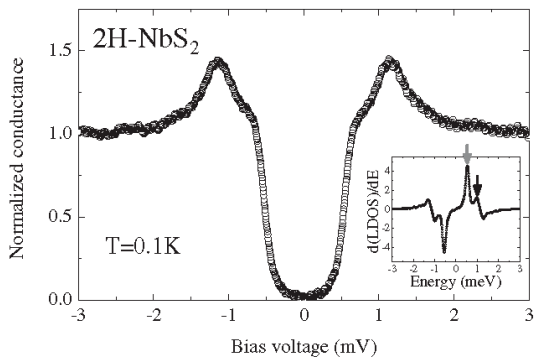}
\caption{STM measurements of the conductance versus V in mV of 2$H$-NbS$_2$, showing strong evidence for two superconducting gaps. The inset is the derivative of the conductance in meV.  Reprinted with permission of I. Guillam{\'o}n, H. Suderow, S. Vieira, L. Cario, P. Diener,  P. Rodi{\`e}re. Superconducting density of states and vortex cores of $2H$-NbS$_2$.  Phys. Rev. Lett.  101 (2008) 166407.  Copyright \copyright2008, American Physical Society.}}
\label{fig5}
\end{figure}
In $2H$-TaSe$_2$, an incommensurate CDW forms below 125 K, and the CDW becomes commensurate with the lattice below 90K, but $2H$-TaSe$_2$ also becomes superconducting at $T_{\rm c}=0.2$ K, below which the commensurate CDW and the superconductivity coexist \cite{WDM,Liu,Inosov}.

One of the most unusual of the $2H$-TMDs is $2H$-NbS$_2$\cite{NaitoTanaka,Molinie,Kennedy,Onabe,Pfalzgraf,Hamaue,Guillamon,Tissen}, which has $T_c=5.84$ K \cite{Onabe}.  Most striking is that there is no evidence for any CDWs.    The Hall constant is positive and nearly independent of $T$ for $T>T_c$, and  there are no anomalies in $\rho_{||}(T)$ and $\rho_{\perp}(T)$  \cite{NaitoTanaka}.  But, since there are two Fermi surfaces, neither one of which is gapped by CDWs,  there might be different superconducting gaps on these different Fermi surfaces.  Recently, scanning tunneling microscopy (STM) experiments were performed on $2H$-NbS$_2$ by Guillam{\'o}n {\it et al.}. The authors  did not observe the star-shaped  CDW that was found by STM in 2$H$-NbSe$_2$ \cite{Hess2}.  However, they did observe two  superconducting gaps 0.53 and 0.97 meV in magnitude, as shown in Fig. 5.  This behavior is similar to that seen in MgB$_2$ and in  pnictide superconductors \cite{Klemm1}.  $T_{\rm c}$ in $2H$-NbS$_2$ was raised from 6 K to 8.9 K by applying 20 GPa of hydrostatic pressure \cite{Tissen}.

\begin{figure}
\center{\includegraphics[width=0.25\textwidth]{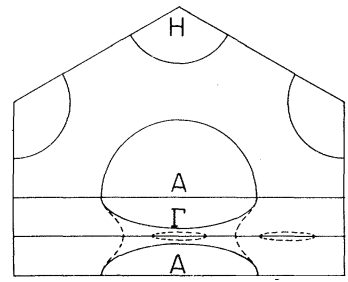}
\caption{Fermi surface of 2$H$-NbSe$_2$ calculated by L. F. Mattheiss. Band structure of transition-metal dichalcogenide layer compounds.  {\it Phys. Rev. B} {\bf 8}, 3719 (1973) (Solid curves), and proposed modifications due to the CDW (dashed curves).  Translation of the pancake at $\Gamma$ by ${\bm Q}_{\rm CDW}$ is also shown.  Preprinted with permission of J. E. Graebner and M. Robbins.  Fermi-surface measurement in normal and superconducting 2$H$-NbSe$_2$. {\it Phys. Rev. Lett.} {\bf 36}, 422 (1976).  Copyright $\copyright$1976, American Physical Society.}\label{fig6}}
\end{figure}

The most studied TMD is 2$H$-NbSe$_2$ \cite{WDM,Nohara,NaitoTanaka,Tonjes,Inosov,Hess2,Revolinsky,Garoche1,Oglesby,Yaron,Koorevaar,Frindt,FonerMcNiff,deTrey,Jerome,Frindt2,Schwall,Sambongi,Jones,Bachmann,Beal1,BevoloShanks,Noto,Smith1,Chu,Molinie,Obolenskii1,Obolenskii2,Obolenskii3,Hess1,Graebner,LeeHNS,EdwardsFrindt,ClaymanFrindt,Clayman,Kennedy,LeeDH,Toyota,Klose,Kobayashi,Muto,Monceau1,MorrisColeman1, MorrisColeman2,MorrisColeman3,Morris,Coleman1,Prober1,Prober2,Dalrymple,Denhoff,Wada1a,Wada1b,Ghoshray,Skripov,Borisenko,Mattheiss,Suderow,Dynes}.
The best samples were made by Oglesby {\it et al.} \cite{Oglesby}, which were large enough for inelastic neutron scattering studies of the vortex lattice \cite{Yaron}.  But in comparison with the other TMDs, the most interesting facts about it are: (1) it has the highest $T_{\rm c}$ value, 7.2 K, of any pristine TMD, (2) There is an incommensurate CDW below $T_{\rm CDW}=35$ K which could consist of one or two components, in addition to a superconducting gap, and (3) there may be  pseudogaps associated with these one or possibly two CDW components that extend very far above $T_{\rm CDW}$.  Direct evidence for the CDW coexisting in the superconducting state was presented by Hess \cite{Hess1,Hess2}. The first magnetothermal oscillation evidence of the Fermi surface changes due to the CDW was presented by Graebner and Robbins \cite{Graebner}, and their proposed Fermi surface distortion is shown in Fig. 6.  Possible ARPES evidence for the two CDW gaps and their pseudogap features well above $T_{\rm CDW}$ was first presented by Borisenko {\it et al.}\cite{Borisenko}. This behavior is strikingly similar to the pseudogap behavior commonly seen in the cuprates \cite{Klemm2}.  However, it should be remembered that no evidence for any magnetic ordering has been presented in any of the many studies of $2H$-NbSe$_2$, only a very small selection of which have been cited here. The superconductivity is not particularly anisotropic, as $\kappa_{\perp}$ and $\kappa_{||}$ are about 13.5 and 54 \cite{Klemm1,Schwall,Toyota,Dalrymple}. In addition, application of hydroscopic pressure  raises $T_{\rm c}$ up to about 8.0 K \cite{Jerome,Smith1,Sambongi,Chu,Molinie,Obolenskii1}, and it suppresses the CDWs, as $T_{\rm CDW}$ decreases linearly with increasing $P$ \cite{Chu}.  Uniaxial pressure along the $c$-axis decreases the resistivity anisotropy to a factor of 5, and both $\rho_{||}(P)$ and $\rho_{\perp}(P)$ exhibit knees at about 30 kbar \cite{Frindt2}. $^{93}$Nb Knight shift measurements on $2H$-NbSe$_2$ were hampered by broad lines at low $T$\cite{Wada1a}, but $^{77}$Se Knight shift measurements yielded $^{77}K(T)$ consistent with the standard model of the electron spin susceptibility, indicating spin-singlet superconducting pairing in $2H$-NbSe$_2$\cite{Wada1b}.  More recently, the CDW state above $T_{\rm c}$ was probed by $^{93}$Nb and $^{77}$Se Knight shift measurements in $2H$-NbSe$_2$\cite{Ghoshray,Skripov}.  $^{93}K(T)$ was strongly increasing for $T<T_{\rm CDW}$\cite{Ghoshray}, opposite to the usual Yosida behavior.  The amplitude of the CDW could be measured from the line width of the $^{77}$Se signal\cite{Skripov}, and fit a Bardeen, Cooper, Schrieffer (BCS) $T$ dependence.

\begin{figure}
\center{\includegraphics[width=0.4\textwidth]{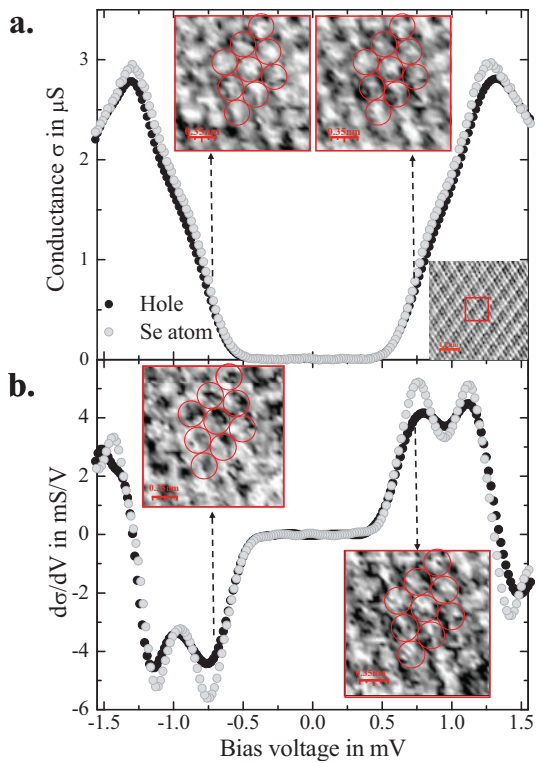}
\caption{(color online) (a) Conductance $\sigma(V)$ curves on 2$H$-NbSe$_2$ at 100 mK on top of Se atoms and in between them. Lower right inset: a typical topographical image of the Se atoms and the CDW on top of it. (b) $d\sigma/dV$ versus $V$.  Red circles indicate the Se positions. Reprinted with permission of I. Guillam{\'o}n, H. Suderow, F. Guinea, and S. Vieira. Intrinsic atomic-scale modulations of the superconducting gap of 2$H$-NbSe$_2$.  {\it Phys. Rev. B} {\bf 77}, 134505 (2008). Copyright\copyright 2008, American Physical Society.}\label{fig7}}
\end{figure}

The first STM evidence of a star-shaped CDW pattern in superconducting 2$H$-NbSe$_2$ was presented by Hess {\it et al.}\cite{Hess2}.  Although there have been many STM studies of 2$H$-NbSe$_2$ since then, a notable one by Guillam{\'o}n {\it et al.} observed the SC gap anisotropy from 0.6 to 1.4 meV.  In addition, the conductance curves showeds locking of the SC order parameter onto the six-fold planar anisotropic CDW order parameter at 0.75 and 1.2 mV, as shown in Fig. 7. This is extremely unusual, and indicates the intricate interplay between these two orderings. Hence, the SC order parameter is highly anisotropic, and it appears that this anisotropy may be directly related to the preexisting six-fold star-shaped CDW anisotropy, which competes for the same Fermi surface with the SC order parameter. But the most direct evidence for an overall $s$-wave orbital symmetry of the superconducting order parameter was obtained by STM with a superconducting tip\cite{Dynes}, as shown in Fig. 8.  These authors succeeded in forming a Josephson junction between the Pb/Ag STM tip and the NbSe$_2$, but noted that the gap anisotropy they measured with a non-superconducting tip, from 0.7 to 1.4 meV, could not account for the low value of $I_{\rm c}R_{\rm n}$ they obtained.  This value was about 20\% lower than that calculated from the minimum NbSe$_2$ gap, suggesting that the presence of the CDW could diminish the value from the anticipated Ambegaokar-Baratoff result for two isotropic superconductors in the absence of a CDW.  Those authors noted that this was similar to the results obtained in the cuprates.\cite{Dynes}

\begin{figure}
\center{\includegraphics[width=0.25\textwidth]{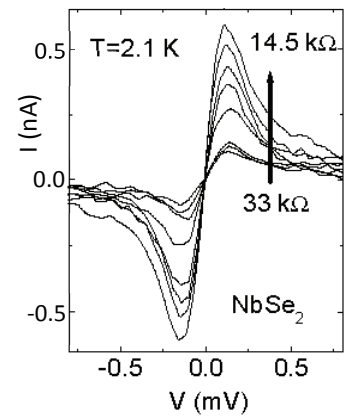}
\caption{Josephson pair current between the superconducting STM Pb/Ag tip and 2$H$-NbSe$_2$ at 2.1 K and junction resistances ranging from 33 k$\Omega$ to 14.5 k$\Omega$, as indicated by the arrow. Reprinted with permission of O. Naaman, R. C. Dynes, and E. Bucher.  Josephson effect in Pb/I/NbSe$_2$ scanning tunneling microscope junctions.  Int. J. Mod. Phys. B {\bf 17}, 3569 (2003). Copyright\copyright World Scientific Publishing Company.}}\label{fig8}
\end{figure}

Resistivity  measurements on 4$H$(b)-TaS$_2$ \cite{DiSalvo3}, with alternating layers of octahedral and trigonal prismatic T$\chi_2$ configurations, were found to exhibit strong evidence for two non-superconducting transitions, with  finite discontinuous drops in $\rho_{||}(T)$ at 315 and 21 K, well above  the superconducting transition at $T_{\rm c} = 2.1$ K \cite{Wattamaniuk}.  In six of eight samples studied, $\rho_{\perp}(T)$ increases monotonically  with decreasing $T$, without any hint of anomalies at 315 and 22 K, until it drops suddenly to zero at $T_{\rm c}$.  The similarity with the behavior seen in Fig. 2 for $2H$-TaS$_2$ strongly suggests that these higher transitions are likely to be CDW transitions.  Under hydrostatic pressure,  the 22 K CDW $\rho_{||}$ anomaly is suppressed by 8.5 kbar, the 315 K CDW $\rho_{||}$ anomaly is suppressed by 35 kbar, and $T_{\rm c}$ increases to 4.35 K by 28 kbar. \cite{Friend}.  De Haas Shubnikov oscillations indicated that the Fermi surface of 4$H$(b)-TaS$_2$ is very 2D, or cylindrical \cite{Fleming}.   The anisotropy  of the upper critical field  of $4H$(b)-Ta$_{0.8}$Nb$_{0.2}$Se$_2$ is weak and nearly independent of temperature \cite{Ikebe1}.

\section{Intercalated transition metal dichalcogenides}
The first example of intercalation of TMD's with organic molecules was made by Weiss and Ruthardt using 2$H$-TiS$_2$ and aliphatic amides \cite{Weiss}.  Trying a similar reaction on a superconducting TMD,
Gamble {\it et al.} (including this author) found that by immersing crystalline $2H$-TaS$_2$ in liquid pyridine at 200$^{\circ}$ C, the TMD crystals grew visibly, and the nominal composition of the intercalated TMD was found to be TaS$_2$(pyridine)$_{1/2}$ \cite{GDKG}, with a $T_c$ that had increased upon intercalation to 3.5 K from 0.6 K in pristine $2H$-TaS$_2$.  Subsequently, it was found that a large class of Lewis bases could intercalate into a variety of TMDs \cite{Gamble,Meyer}.  This class included octadecylamine, which when intercalated into TaS$_2$, led to a $c$-axis repeat distance of 5.7 nm.  However, $T_c$ did not change significantly upon the length of the carbon chain of aliphatic amine intercalants \cite{Gamble,DiSalvo1}.  Intercalation of paramagnetic cobaltocene  into both $2H$-TaS$_2$ and  SnSe$_2$ led to superconducting compounds \cite{GambleThompson,Formstone}.  Nearly all organic or organometallic intercalation compounds of $2H$-TaS$_2$ had onset $T_c$ values close to 3.5 K.     The increase in $T_c$ upon intercalation of $2H$-TaS$_2$ with Lewis bases was found to be due to the suppression of the CDWs by the intercalation process in that material \cite{DiSalvo2,DiSalvo4}. Curiously, when $2H$-NbS$_2$ was intercalated with pyridine and other organic Lewis bases, $T_c$ decreased from its pristine value of 5.8-6.3 K to 3-4 K, nearly the same as for the intercalation compounds of $2H$-TaS$_2$. Since $2H$-NbS$_2$ does not give rise to CDW formation, but instead has different superconducting gaps on the two major Fermi surfaces \cite{Guillamon}, one might suppose that one of these superconducting gaps might have been at least partially suppressed upon intercalation.  However, there is to date no direct evidence of this.

\begin{figure}
\center{\includegraphics[width=0.35\textwidth]{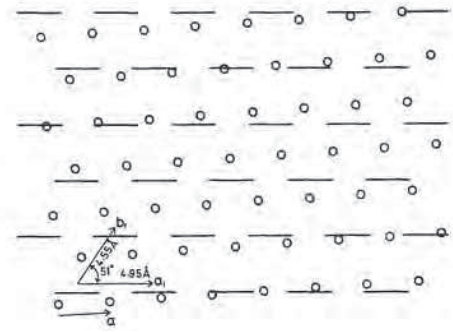}
\caption{Sketch of one of the three domains of the incommensurate pyridine structure  in the nominally single-phase $c/2=1.203$ nm composition TaS$_2$(pyridine)$_{0.54}$ \cite{Kashihara2}. The circles and line segments respectively represent  $S$ ions and pyridine molecules lying perpendicular to the TaS$_2$ layers, which have shifted during intercalation from the $2H$ structure so that the $S$ ions on adjacent layers are vertically on top of one another.  See text.  Reprinted with permission of Y. Kashihara, H. Yoshioka.  Electron dirrfaction study of TaS$_2$ intercalated with pyridine. J. Phys. Soc. Jpn. 50 (1981) 2084.  Copyright \copyright1981, Physical Society of Japan.}}\label{fig9}
\end{figure}

Since it was clear that such intercalation chemistry involved charge transfer from the intercalant to the host TMD,  Gamble {\it et al.} first imagined that a possible structure of TaS$_2$(pyridine)$_{1/2}$ would involve the unpaired electrons on the nitrogen atoms in the pyridine molecules being located as close as possible to the host TaS$_2$ layers \cite{Klemm1,GDKG}.  In the case of TaS$_2$(pyridine)$_{1/2}$, the charge of about one electron for two pyridine molecules was determined to be transferred \cite{Beal2,Ehrenfreund}, leading to the possible formation of bipyridyl within the intercalant layers \cite{Schollhorn}. The charge transfer was found to be facilitated by the presence of water, which can produce pyridinium (H$_2$O)$^{+}$ ions \cite{Johnson}.

Intercalation of $2H$-TaS$_2$ with pyridine was found to be much more complicated than originally imagined by the author and his co-workers.  Electron microscopy studies revealed screw dislocations in the host TaS$_2$ layers, due to the difficulty in making single-phase $2H$-TaS$_2$ without annealing it from $4H$(b)-TaS$_2$ \cite{DiSalvo3,FernandezMoran}.  These screw dislocations cause the $2H$-TaS$_2$ layers to exfoliate upon intercalation.  In addition, the original intercalation procedure led to a mixed composition of nominal TaS$_2$(pyridine)$_{1/2}$, with the different $c$-axis values $c/2=1.203$ and 1.193 nm, respectively \cite{Thompson3}.  By dissolving excess sulfur in the pyridine, Thompson found that the ``single phase'' intercalation compound with only $c/2=1.203$ nm and a greatly reduced tendency for exfoliation upon intercalation was obtained \cite{Thompson4}.  Neutron scattering, electron diffraction, and nuclear magnetic resonance (NMR) studies of NbS$_2$(pyridine)$_{1/2}$ and TaS$_2$(pyridine)$_{1/2}$ indicated that the nitrogen atoms in the intercalant pyridine molecules were midway between the TMD layers, so that the planes of the pyridine molecules were perpendicular to the TaS$_2$ layers \cite{Parry,Riekel,McDaniel,Kashihara1,Kashihara2}.  The mixed phase $c/2=1.203,1.193$ nm material obtained by immersion the TMD in pure pyridine involved a horizontal shift of the TMD layers, so that the S atoms lay directly above one another after intercalation, which was not the case in the pristine $2H(a)$ polytype \cite{Parry}.  Second, Parry {\it et al.} found that this procedure led to a pyridine superlattice that was commensurate with the TaS$_2$ lattice on both sides of each pyridine layer, with the superlattice cell dimensions of 13$a\times2\sqrt{3}a$ in terms of the $a$ hexagonal planar lattice constant of $2H$-TaS$_2$, and actual composition TaS$_2$(pyridine)$_{6/13}$ \cite{Parry}.  However, Kashiwaya and Yoshioka found that in addition to this superlattice, there was a second superlattice of cell dimensions $9a\times2\sqrt{3}a$ and composition TaS$_2$(pyridine)$_{4/9}$, and the two superlattices of pyridine-intercalated TaS$_2$ formed from pure pyridine contained domains of both commensurate pyridine superlattices \cite{Kashihara2}.

Kashihara and Yoshioka also studied a single crystal of the  ``single-phase'' $c/2=1.203$ nm by electron diffraction \cite{Kashihara2}.  They found that the pyridine molecules lie perpendicular to the TaS$_2$ layers, as in the previous sample, but formed one-dimensional chains that are incommensurate with the TaS$_2$ layers in both directions.  The pyridine lattice of this incommensurate TaS$_2$(pyridine)$_{0.54}$ structure has constants $a_1=0.495$ nm and $b_1=0.455$ nm, with the angle $\alpha=51^{\circ}$ between those lattice vectors. In addition, this incommensurate pyridine lattice is also inclined by 3.3$^{\circ}$ relative to the TaS$_2$ lattice. Moreover, there are actually three domains of this incommensurate lattice within a single crystal, two of which are rotated by $\pm120^{\circ}$ from that description \cite{Kashihara2}.  The first domain of this incommensurate lattice is sketched in Fig. 9.

The normal state properties of  TaS$_2$(pyridine)$_{1/2}$ are very different from those of pristine 2$H$-TaS$_2$ crystals\cite{TGK}. The magnetic susceptibility $\chi_{\perp}(T)$ for TaS$_2$(pyridine)$_{1/2}$ is a constant above $T_c$, due to the destruction of the CDWs upon intercalation, and does not exhibit any anomaly at or below the $T_{\rm CDW}=75$ K of pristine 2$H$-TaS$_2$\cite{TGK}.   $\rho_{||}(T)$  for TaS$_2$(pyridine)$_{1/2}$ is rather similar to that of pristine $H$-TaS$_2$ above $T_{\rm CDW}=75$ K, but the anomaly present in Fig. 3 for pristine 2$H$-TaS$_2$ at $T_{\rm CDW}=75$ K is completely absent, and $\rho_{||}(T)$ decreases with $T$ in a quasi-linear fashion at lower $T$.  The values of $\rho_{||}(300$ K) are nearly the same in the pristine and intercalated materials. On the other hand, $\rho_{\perp}(T)$ for TaS$_2$(pyridine)$_{1/2}$ is enormously different from that pictured for pristine 2$H$-TaS$_2$ in Fig. 3.  The minimum in the $\rho_{\perp}(T)$ curve for $2H$-TaS$_2$ is missing in the corresponding curve for TaS$_2$(pyridine)$_{1/2}$.  Moreover, $\rho_{\perp}(300$ K) increases by more than four orders of magnitude upon intercalation.  For 2$H$-TaS$_2$, the values measured were $\rho_{||}(300\>{\rm K})=1.5\mu\Omega$m and $\rho_{\perp}(300\>{\rm K})=22\mu\Omega$m.  For TaS$_2$(pyridine)$_{1/2}$, $\rho_{||}(300\>{\rm K})=3\mu\Omega$m and $\rho_{\perp}(300\>{\rm K})=0.20\Omega$m \cite{TGK}.  This room temperature resistivity anisotropy of about $10^5$ for TaS$_2$(pyridine)$_{1/2}$ is almost certainly limited by electrical shorts due to screw dislocations and exfoliations, so that the actual resistivity anisotropy is likely to be much larger than this.  Proton spin-lattice relaxation rates in TaS$_2$(pyridine)$_{1/2}$ and in NbS$_2$(pyridine)$_{1/2}$ were measured \cite{Wada2,Wada3,Wada4}. These results were later found to be similar in their $T$ dependencies to the corresponding $1/T_1$ rates measured in high-$T_{\rm c}$ superconductors.  Most important,  the doubly-incommensurate pyridine superlattice domain structures relative to the TaS$_2$ layers imply that the $c$-axis normal state electronic transport is completely incoherent, as in the high-$T_c$ superconductor Bi$_2$Sr$_2$CaCu$_2$O$_{8+\delta}$ and in the organic layered superconductor $\kappa$-(ET)$_2$Cu[N(CN)$_2$]Br, where ET is bis(ethylenedithio)tetrathiafulvalene.

\begin{figure}
\center{\includegraphics[width=0.36\textwidth]{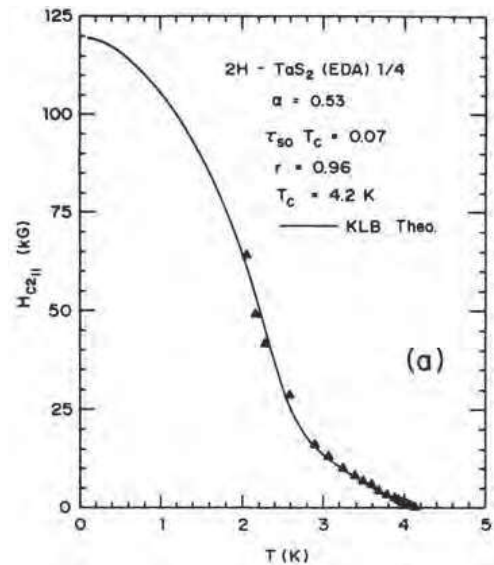}
\caption{Fits of $H_{c2,||}(T)$ for TaS$_2$(ethylenediamine)$_{1/4}$.  The solid line is a fit to the Klemm, Luther, Beasley theory using $r=0.96$, $T_{\rm c}=4.2$ K, $\alpha=0.53$, and $\tau_{so}k_BT_c/\hbar=0.105$\cite{KLB}). Reprinted with permission of R.V. Coleman, G.K. Eiserman, S.J. Hillenius, A.T. Mitchel, J.L. Vicent.  Dimensional crossover in the superconducting intercalated layered compound 2$H$-TaS$_2$.  Phys. Rev. B  27 (1983) 125.  Copyright \copyright1983, American Physical Society.}}\label{fig10}
\end{figure}

The upper critical field $H_{c2}(T)$ of several intercalation compounds of $2H$-TaS$_2$ and nitrogen containing organic molecules were made by Prober {\it et al.} and by Coleman {\it et al.}, and the results  of $H_{c2,||}(T)$ for TaS$_2$(ethylenediamine)$_{1/4}$ are shown in Fig. 10 \cite{Prober2,Coleman}.    Dimensional crossover is signaled by the upward curvature below $T_c$ for $H_{c2,||}(T)$.  Stronger evidence for dimensional crossover was provided in detailed fits of $H_{c2,||}(T)$ for TaS$_2$(pyridine)$_{1/2}$, TaS$_2$(aniline)$_{3/4}$ and TaS$_{0.4}$Se$_{1.6}$(collidine)$_{1/6}$ to the Klemm, Luther, Beasley (KLB) theory \cite{Prober2,KLB}.  Such fits of $H_{c2,||}(T)$ for  TaS$_2$(ethylenediamine)$_{1/4}$ to the KLB theory were made by Coleman {\it et al.} \cite{Coleman,KLB}, and the results are shown in Fig. 10. The fitting parameters are the slope $\alpha$ of $H_{c2,||}(T)$ at $T_{\rm c}$ relative  to the Pauli limit, the dimensional-crossover parameter $r$, which is essentially the square of the ratio of the $T=0$ superconducting coherence length $\xi_c(0)$ in the $c$-axis direction to the $c$-axis repeat distance $s$, and the spin-orbit scattering rate, which probably arises due to multiple scatterings off the interfaces between the TaS$_2$ layer surfaces containing the heavy element Ta (Z=73) and the light organic layers. Surface spin-orbit scattering has been shown to follow the Abrikosov-Gor'kov value $\propto Z^4$\cite{Klemm1}.  In this case, the evidence for dimensional crossover effects is strong.\cite{Coleman,KLB}  In those measurements, $H_{c2,||}(0)$ violates the Pauli limit of $\mu_0H_{c2}(0)=1.85 T_c$(T/K), which is $\mu_0H_{c2}(0)=7.7$ T for $T_{\rm c}=4.2 $K, by about a factor of 1.6 \cite{Klemm1}.  In addition, similar studies of these and Ta$_{1-x}$Nb$_x$S$_2$(pyridine)$_{1/2}$ were made \cite{Ikebe2,Ikebe3}, and were found to violate the Pauli limit by about a factor of 2.

The superconducting misfit layer compounds comprise another very interesting class of TMD intercalation compounds \cite{Wiegers1,Wiegers2,Nader,Reefman,Wulff1,Wulff2,Smontara,Roesky,Monceau2,Kacmarcik1,Szabo,Samuely,Kacmarcik2}.  These materials have compositions $(M\chi)_{1+x}(T\chi_2)_m$, where $M$ is  either a transition metal or a lanthanide, $0<x<1$, $m=1,2,3$, and $T$ is a group V transition metal.  These materials with $m=1$ consist of alternating layers of the hexagonal $T\chi_2$ layers and the $M\chi$ layers, which can either be hexagonal with a different lattice constant, or can even have a different planar lattice structure.  In the most interesting case of $M=$ La and $T=$ Nb with $\chi=$ Se, the tetragonal insulating LaSe layers with $x=0.14$ are intercalation layers of the hexagonal superconducting NbSe$_2$ single layers \cite{Kacmarcik1,Szabo,Samuely,Kacmarcik2}.  In this case, upper critical field studies shown in Fig. 11  demonstrate that the material undergoes dimensional crossover from 3D to 2D behavior at about $T^{*}=1.1$ K \cite{KLB}, just below $T_c=1.2$ K determined resistively.   However, $H_{c2,||}(0)$ violates the conventional Pauli limit (2.2 T) by the enormous factor of 8.4. This is almost as unconventional as for the ferromagnetic superconductors URhGe and UCoGe, the upper critical fields of which violate the Pauli limit by at least a factor of 20.  However,  in the insulating LaSe layers have only the slightly larger $Z=57$  than do the  Nb ions, $Z=41$. The relevant spin orbit scattering time must satisfy $\tau_{\rm so}k_BT_{\rm c}/\hbar < 10^{-3}$\cite{KLB},  at least an order or magnitude less than for the TaS$_2$(EDA)$_{1/4}$ $H_{c2,||}(T)$ fit pictured in Fig. 10. This might raise a serious issue as to whether (LaSe)$_{1.14}$(NbSe$_2$) were a singlet superconductor. But, strong evidence for that proposition is presented in the following. At $T=0.58$ K, the measured $B_{c2}(\theta)$ fits the Tinkham thin film formula \cite{Klemm1,Tinkham}, as shown in Fig. 11, further demonstrating the 2D superconductivity well below $T^{*}$.
 \begin{figure}
\center{\includegraphics[width=0.45\textwidth]{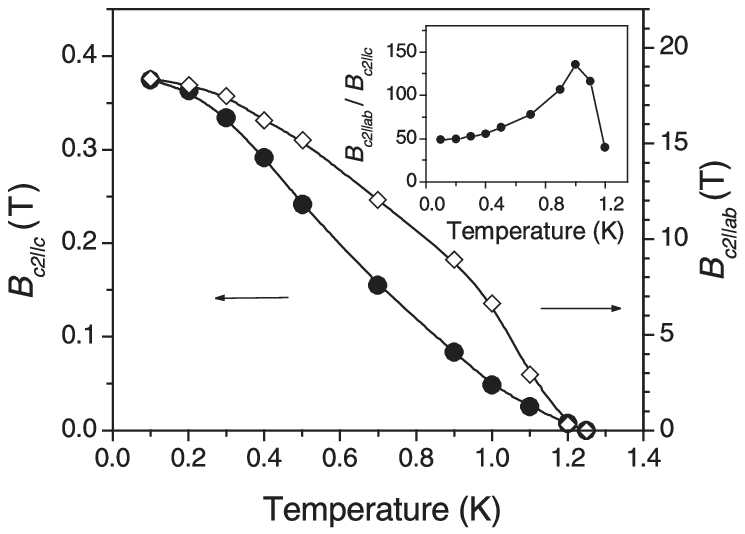}}
\center{\includegraphics[width=0.35\textwidth]{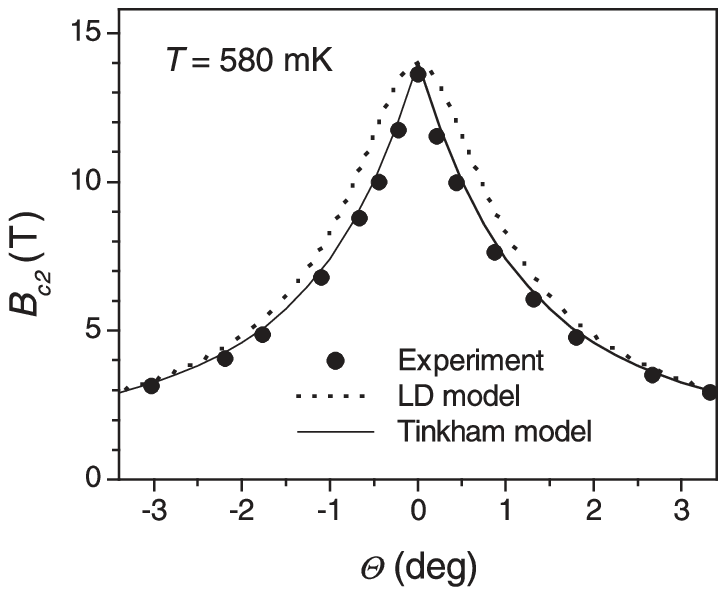}
\caption{Upper: $B_{c2,\perp}(T)$ (left scale) and $B_{c2,||}(T)$ (right scale) of (LaSe)$_{1.14}$(NbSe$_2$). Inset: anisotropy $B_{c2,||}(T)/B_{c2,\perp}(T)$.  Lower: $B_{c2}(\theta)$ at $T=0.58$ K of (LaSe)$_{1.14}$(NbSe$_2$).  The solid and dotted curves represent the Tinkham formula for 2D thin films and the 3D anisotropic mass model \cite{Tinkham,Klemm1}. Reprinted with permission of P. Samuely, P. Szab{\'o}, J. Ka{\v c}mar{\v c}ik, A.G.M. Jansen, A. Lafond, A. Meerschaut, A. Briggs.  Two-dimensional behavior of the naturally layered superconductor (LaSe)$_{1.14}$(NbSe$_2$). Physica C 369 (2002) 61.  Copyright \copyright2002, Elsevier.}}\label{fig11}
\end{figure}
\begin{figure}
\center{\includegraphics[width=0.4\textwidth]{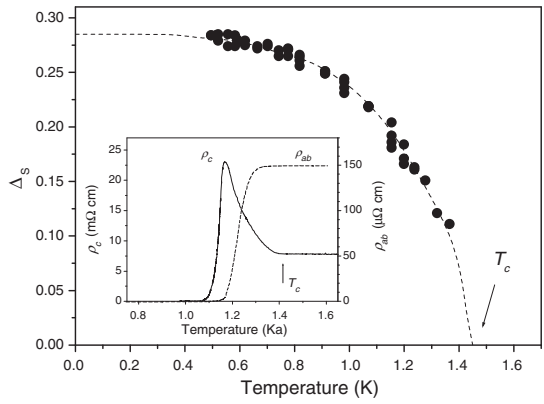}
\caption{$\Delta_S(T)$ in meV measured by STM on (LaSe)$_{1.14}$(NbSe$_2$). The dashed curve is a extrapolation of $\Delta_S(T)\rightarrow0$ using the BCS theory. Inset: $\rho_c(T)$ (left scale) and $\rho_{ab}(T)$ (right scale) vs. $T$ in K.  The vertical bar labeled $T_c$ indicates the point at which $\Delta_S(T)\rightarrow0$.  Reprinted with permission of  J. Ka{\v c}mar{\v c}ik, P. Szab{\'o}, J.G. Rodrigo, H. Suderow, S. Vieira, A. Lafond, A. Meerschaut.   Intrinsic Josephson junction behaviour of the low $T_c$ superconductor (LaSe)$_{1.14}$(NbSe$_2$). Physica C 468 (2008) 543.  Copyright \copyright2008, Elsevier.}}\label{fig12}
\end{figure}

  The insulating organic layers in intercalated TMDs and the insulating LaSe layers in (LaSe)$_{1.14}$(NbSe$_2$) show convincing evidence of behaving as intrinsic Josephson junctions, just as do the insulating pyridine layers in TaS$_2$(pyridine)$_{1/2}$, the Cu[N(CN)$_2$]Br layers in $\kappa$-(ET)$_2$Cu[N(CN)$_2$]Br, and  the BiO double layers in  Bi$_2$Sr$_2$CaCu$_2$O$_{8+\delta}$\cite{WKK}.  STM measurements of two different cleaves confirmed that the LaSe layers are tetragonal and the NbSe$_2$ layers are hexagonal\cite{Kacmarcik2}, so that the lattice periodicities are completely incommensurate in both planar directions.  Pb STM measurements of the superconducting gap in (LaSe)$_{1.14}$(NbSe$_2$ provided strong evidence for an isotropic, strong-coupling, superconducting gap $2\Delta_S(0)=4.5k_BT_c$, where $T_c=1.42$ K is obtained by extrapolating the BCS $T$ dependence of $\Delta_S(T)$ to zero \cite{Kacmarcik2}, as shown in the lower panel of Fig. 12.  It is very mysterious as to why a BCS superconductor would have an $H_{c2,||}(0)$ in violation of the Pauli limit by nearly an order of magnitude, especially with the largest elemental $Z=57$.  This is indeed ``unconventional'' behavior, and is much larger than for any other TMD and even than for the most anomalously large Pauli-limit violating organic layered superconductor, $\kappa$-(ET)$_4$Hg$_{2.89}$Br$_8$, for which the violation is only by a factor of 3, which might be explainable with Hg having $Z=80$\cite{Klemm1}.  Moreover, in the inset of Fig. 12, $\rho_c(T)$ exhibits a strong increase  below the onset of the superconducting $\Delta_{\rm S}(T)\ne0$, which is very similar to that observed in underdoped Bi$_2$Sr$_2$CaCu$_2$O$_{8+\delta}$,\cite{WKK} signifying strongly incoherent $c$-axis quasiparticle tunneling and intrinsic Josephson junction behavior.

In addition to the spectacular 2D behavior of (LaSe)$_{1.14}$(NbSe$_2$) shown in Figs. 10, 11, and 12, the misfit compound (SnS)$_{1.17}$(NbS$_2$) was made, and the anisotropy of $H_{c2}(\theta,T)$ was measured \cite{Nader}. At 500 mK,  $H_{c2}(\theta)$ fit the anisotropic mass model\cite{KLB,Klemm1}, the form of the dotted curve in the lower panel of Fig. 11, signifying 3D behavior.  No evidence for dimensional crossover was present in $H_{c2,||}(T)$.

 A large number of layered superconductors were made by doping (or intercalating)   $2H$-MoS$_2$, $2H$-TaS$_2$, $1T$-TaS$_2$, $4H$(b)-TaS$_2$, $2H$-ZrS$_2$, and $2H$-ZrSe$_2$ with alkali \cite{Levy,Somoano,Somoanoreview,Woolam,Fang,Nozuyama,Ahmad,Onuki}.  The resulting intercalation compounds were very air-sensitive and rather 3D in their properties.  Recently, Ye {\it et al.} showed that it was possible to induce superconductivity into 2$H$(b)-MoS$_2$ up to $T_{\rm c}$~$\sim$~11~K by electrostatic gating\cite{Ye}, and those results are combined with chemical doping $T_{\rm c}$ results in Fig. 13.
 \begin{figure}
\center{\includegraphics[width=0.45\textwidth]{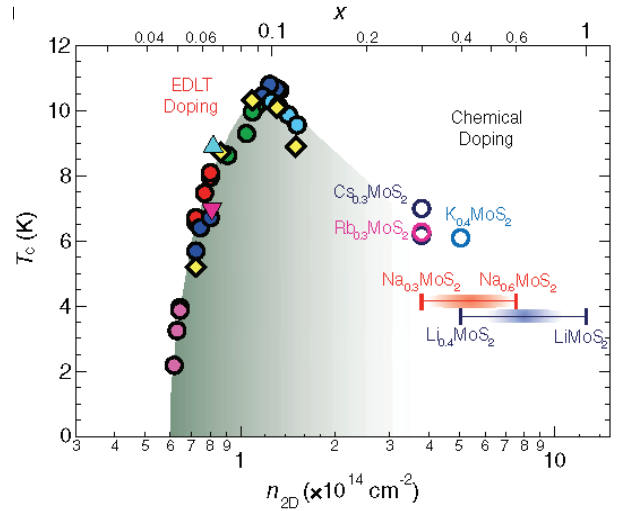}
\caption{(Color online) Unified phase diagram of superconducting $T_c$ from both electrostatically (filled circles) and chemically (open circles) doped 2$H$-MoS$_2$ versus doping concentration $x$ (upper axis) and carrier density (lower axis).  Reprinted with permission of J. T. Ye, Y. J. Zhang, R. Akashi, M. S. Bahramy, R. Arita, and Y. Iwasa.  Superconducting dome in a gate-tuned band insulator.  {\it Science} {\bf 338}, 1193 (2012).  Copyright\copyright2012 American Association for the Advancement of Science.}\label{fig13}}
\end{figure}

\begin{figure}
\center{\includegraphics[width=0.45\textwidth]{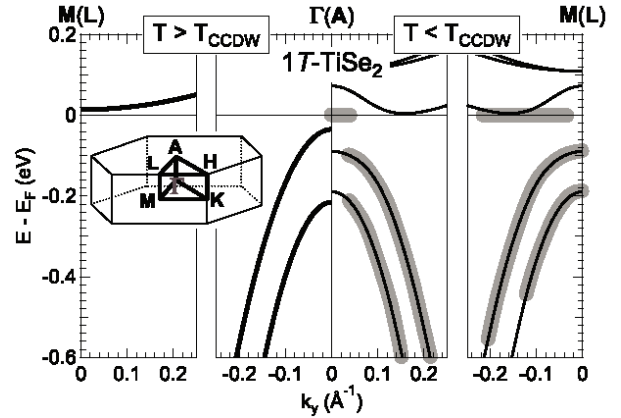}
\caption{Schematic electronic structures of 1$T$-TiSe$_2$ in the normal (left) and commensurate CDW (right) phases. Band structure features that carry significant weight are highlighted in thick gray curves. Inset:  Bulk Brillouin zone. Reprinted with permission of K. Rossnagel.  On the origin of charge-density waves in select transition metal dichalcogenides.  J. Phs.: Condens. Matter {\bf 23}, 213001 (2011). Copyright \copyright2011 IOP Publishing Ltd. }}\label{fig14}
\end{figure}

In addition, intercalation compounds Ta$M_x$S$_2$ of $2H$-TaS$_2$ were made with $x=1$ using transition metals such as Hg, In, Pb and Sn \cite{DiSalvo3,Dijkstra}.  With $M=$ Bi, $x$ compositions $\frac{1}{3}$ and $\frac{2}{3}$ were also made \cite{DiSalvo3}  The specific heat of TaSnS$_2$ was measured \cite{Dijkstra}, confirming that it was superconducting below 2.95 K.  But the more interesting intercalation of a TMD with a transition metal was the intercalation of $1T$-TiSe$_2$ with Cu \cite{Morosan}.  The schematic electronic structure of the pristine 1$T$-TiSe$_2$ above and below the commensurate charge density wave transition $T_{\rm CCDW}$ was semiempirically determined by Rossnagel, and are pictured in Fig. 14.  The phase diagram of the intercalated Cu$_x$TiSe$_2$ materials is shown in Fig. 15\cite{Morosan}. In this work, small amounts of Cu as an intercalant suppress the CDW and the semi-metallic state in pristine TMD, allowing for superconductivity in TiCu$_x$Se$_2$ for $x\ge 0.02$, with a maximum $T_c =4.0$ K at $x=0.08$.  From the conventional  specific heat peak and the measured $H_{c1,\perp}(T)$ and $H_{c2,\perp}(T)$  for $x=0.06$ only differing by about a factor of 11, the data suggest that TiCu$_{0.06}$Se$_2$ is  a rather ordinary anisotropic type-II superconductor \cite{Morosan}.  No measurements of critical field anisotropies were reported.  This enhancement of superconductivity by transition metal intercalation is similar to the much earlier iron intercalation of $2H$-TaS$_2$, which raised $T_c$ from 0.6 K to about 3K in TaFe$_{0.05}$S$_2$\cite{Coleman}.  Although $H_{c2,\perp}(T)$ showed an anomalously large amount of upward curvature, the $H_{c2}$ anisotropy $H_{c2,||}(T)/H_{c2,\perp}(T)$ was nearly the constant value of 35.  The upward curvature therefore observed in $H_{c2,||}(T)$ could not really be associated with dimensional crossover effects \cite{Coleman}.  In addition, $2H$-TaS$_2$ was intercalated with Cu, and TaCu$_{0.03}$S$_2$ was produced and studied by nuclear magnetic resonance \cite{Zhu}.

Another class of TMD intercalation compounds is the hydrates with transition metal ions as intercalants into $2H$-TaS$_2$ \cite{Lerf,Gygax,Biberacher,Johnston,Kanzaki,Schlicht,Sernetz,vonWesendonk}.  These compounds have the general composition $M_x$(H$_2$O)$_yT\chi_2$, where $T$ can be either Ta or Nb, $\chi=$ S, and $M$ can be an alkali or a rare earth ion.  Usually the compounds are made by electrolytic intercalation processes.  The resulting compositions sometimes do not include water, or can involve alkali hydroxides in place of $M$ and H$_2$O. Typical $T_c$ values range from 2.1 to 5.6 K.  Few superconducting experiments were performed, other than to measure $T_{\rm c}$ and possibly some $H_{c2}$ data.  No order parameter symmetry experiments were performed.  However, since these are non-stoichiometric compounds,  the $c$-axis tunneling through the ion-water layers is most likely incoherent.  These compounds are very similar to the cobaltates such as Na$_x$CoO$_2\cdot y$H$_2$O, which also contain intercalant water layers with dissolved Na ions and remarkably similar $T_{\rm c}$ values and anisotropic normal state and superconducting properties\cite{Klemm1}.

\begin{figure}
\center{\includegraphics[width=0.4\textwidth]{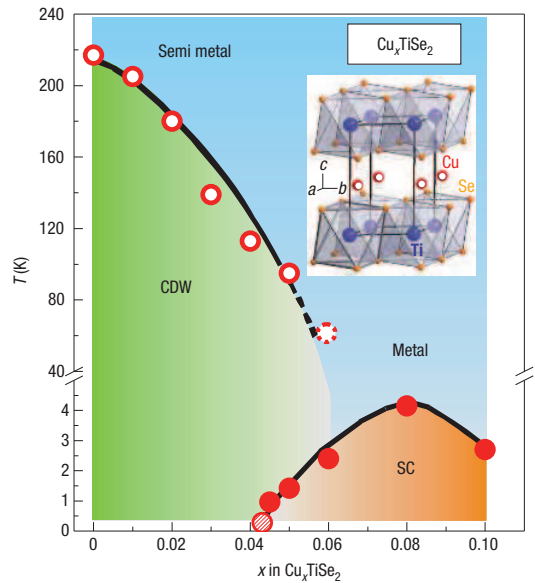}
\caption{(color online) Structure and phase diagram of Cu$_x$TiSe$_2$.  Reprinted with permission of E. Morosan, H. W. Zandbergen, B. S. Dennis, J. W. G. Box, Y. Onose, T. Klimczuk, A. P. Ramirez, N. P. Ong, R. J. Cava.  Superconductivity in Cu$_x$TiSe$_2$.  Nature Phys. {\bf 2}, 544 (2006).  Copyright \copyright2006, Nature Publishing Group.}}\label{fig15}
\end{figure}
\section{Summary and Conclusions}
The transition metal dichalcogenides are very complicated materials, which mimic many of the features seen in the high-$T_c$ cuprates, pnictides, the organic layered superconductors, the cobaltates, and MgB$_2$.  Nearly all of them are complicated by one or more CDWs, which can be either commensurate or incommensurate with the underlying lattice, can exhibit gap nodes (although such nodes are rare), can possibly persist as pseudogaps well above the onset of long-range CDW order, and compete directly with the superconductivity for Fermi surface gapping area below $T_{\rm c}$.   With sufficiently strong applied pressure or by intercalation with a variety of organic or inorganic materials, the CDWs are suppressed, and the superconductivity either appears or is enhanced. $2H$-NbS$_2$ is the exception, as it does not exhibit any CDW formation, but instead has two, rather isotropic, superconducting gaps, presumably on different Fermi surfaces. In this case, intercalation reduces $T_c$, but hydrostatic pressure increases $T_c$, which might be due to  Fermi surface pressure changes.  Scanning tunneling microscopy with and without a superconducting tip provided Josephson junction evidence of a highly anisotropic $s$-wave superconducting order parameter in 2$H$-NbSe$_2$, although the strength of the Josephson coupling was weaker than expected, and attributed to possible interactions with the CDW. Pristine  $2H$-NbSe$_2$ has a complicated star-shaped CDW gap structure that coexists with the superconductivity, and possible pseudogaps well above the onset of CDW order, as in some cuprate superconductors.  The CDW in $2H$-TaS$_2$ has a node on the Fermi surface.  Both the intrinsic misfit compound (LaSe)$_{1.14}$(NbSe$_2$) and a number of  organic intercalates of $2H$-TaS$_2$ and $2H$-NbS$_2$, are not only free of any CDWs, but exhibit strong $H_{c2,||}(T)$ evidence for dimensional crossover from 3D to 2D behavior below $T_c$.  Moreover, $H_{c2,||}(0)$ for (LaSe)$_{1.14}$(NbSe$_2$) appears to violate the ordinary Pauli limit by the factor of 8.4, which is very unconventional, especially since the gap $\Delta(T)$ fits the BCS curve for an isotropic $s$-wave superconductor accurately.  As do the organic intercalates of $2H$-TaS$_2$ and presumably those of $2H$-NbS$_2$, (LaSe)$_{1.14}$(NbSe$_2$) has incoherent $c$-axis transport,  and exhibits dimensional crossover at a temperature below which it behaves as a stack of intrinsic Josephson junctions, precisely as do Bi$_2$Sr$_2$CaCu$_2$O$_{8+\delta}$ and the organic layered superconductor $\kappa$-(ET)$_2$Cu[N(CN)$_2$]Br \cite{Klemm1,PM}.  Thus, the transition metal dichalcogenides and their intercalates, while $s$-wave superconductors, contain many examples that are not at all conventional in their superconducting and normal state properties.

\end{document}